\pdfoutput=1

\documentclass[10pt,letterpaper]{article}

\usepackage[margin=2.5cm]{geometry}
\usepackage[font=sf]{caption}

\usepackage{epsfig}
\usepackage{amssymb}
\usepackage{mciteplus}

\newcommand{\text}[1]{{\mathrm{#1}}}
\newcommand{\un}[1]{\ensuremath{\,\text{#1}}}

\newcommand{\vsd}{\ensuremath{V_{\text{sd}}}}
\newcommand{\vg}{\ensuremath{V_{\text{g}}}}
\newcommand{\vgaceff}{\ensuremath{\vg^\text{ac,eff}}}
\newcommand{\vaac}{\ensuremath{V_\text{RF}}}
\newcommand{\kb}{\ensuremath{k_{\text{B}}}}
\newcommand{\cg}{\ensuremath{C_\text{g}}}
\newcommand{\cgac}{\ensuremath{\cg^\text{ac}}}
\newcommand{\textmu}{\ensuremath{\mu}}
\newcommand{\dvg}{\ensuremath{\delta \vg}}
\newcommand{\vgstar}{\ensuremath{\vg^\text{*}}}

\long\def\symbolfootnote[#1]#2{\begingroup%
\def\thefootnote{\fnsymbol{footnote}}\footnote[#1]{#2}\endgroup} 

\begin{document}

\begin{center}

{\LARGE Carbon nanotubes as ultra-high quality factor mechanical
  resonators} 

\vspace{2em}

A. K. H\"{u}ttel$^{1,*}$\symbolfootnote[0]{
$^*$Present address: Institute for Experimental
  and Applied Physics, University of Regensburg, 93040 Regensburg,
  Germany},
G. A. Steele$^1$, 
B. Witkamp$^1$, M. Poot$^1$,
L. P. Kouwenhoven$^1$ and H. S. J. van der Zant$^1$ 

{\em $^1$Kavli Institute of NanoScience, Delft University of Technology, PO Box 
5046, 2600 GA, Delft, The Netherlands.}

\end{center}

\vspace{0.5em}

\begin{abstract}
  We have observed the transversal vibration mode of suspended carbon
  nanotubes at millikelvin temperatures by measuring the
  single-electron tunneling current. The suspended nanotubes
  are actuated contact-free by the radio frequency electric field of a
  nearby antenna; the mechanical resonance is detected in the
  time-averaged current through the nanotube. Sharp, gate-tuneable resonances
  due to the bending mode of the nanotube are observed, combining resonance frequencies
  of up to $\nu_0=350\un{MHz}$ with quality factors above $Q = 10^5$, much higher than previously reported
  results on suspended carbon nanotube resonators. The measured magnitude
  and temperature dependence of the Q-factor shows a remarkable agreement
  with the intrinsic damping predicted for a suspended carbon nanotube.
  By adjusting the RF power on the antenna, we find that the nanotube
  resonator can easily be driven into the non-linear regime.
\end{abstract}

\vspace{1.5em}

High-quality resonating systems, providing high frequency
resolution and long energy storage time, play an important role
in many fields of physics. In particular in the field of
nanoelectromechanical systems
 \cite{book-cleland,rsci-ekinci:061101}, recent research has led
to the development of high-frequency top-down fabricated mechanical
resonators with high quality factors
 \cite{Naik2006, prl_flowers, nnano_FengRoukes, zolfagharkhani:224101}.
However, when miniaturizing mechanical
resonators to make them lighter and to increase their resonance
frequency  \cite{book-cleland}, the quality factor tends to
decreases significantly from surface effects
 \cite{rsci-ekinci:061101}. High Q-values combined with high
resonance frequencies are an important prerequisite for
applications such as single-atom mass sensing
 \cite{nl-lassagne:3735,nl-chiu:0,nnano-jensen:533} and
fundamental studies of the quantum limit of mechanical motion
 \cite{pt_schwabroukes}. Single-wall carbon nanotubes present a
potentially defect-free nanomechanical system with extraordinary
mechanical properties: in particular the high Young's modulus
($E=1.2\un{TPa}$) in combination with a very low mass density
($\rho=1350\,\text{kg/m}^3$)  \cite{nl-chiu:0,
nature-sazonova:284, nl-witkamp:2904, apl-witkamp:111909}. While
these favorable properties should result in quality factors of
the order of $2\times 10^5$  \cite{PhysRevLett.93.185501}, the
observed Q-factors of nanotube resonators both at room
temperature  \cite{nature-sazonova:284, nl-witkamp:2904,
nl-jensen:3508, nl-eriksson} and in low temperature experiments
 \cite{nl-lassagne:3735,nl-chiu:0} have not exceeded $Q \sim
2000$.

Here we report on the observation of mechanical resonances of a
driven suspended carbon nanotube at low temperatures with quality
factors above $10^5$ and resonance frequencies ranging from 120
MHz to 360 MHz. The resonances are detected with a novel
detection scheme which uses the non-linear gate-dependence of the
current through the suspended nanotube quantum dot. In addition,
we show that the nanotube resonator can easily be tuned to the
non-linear regime, and that the operating temperature affects the
non-linearity and the quality factor of the resonator.

\begin{figure}
\centering
  \epsfig{file=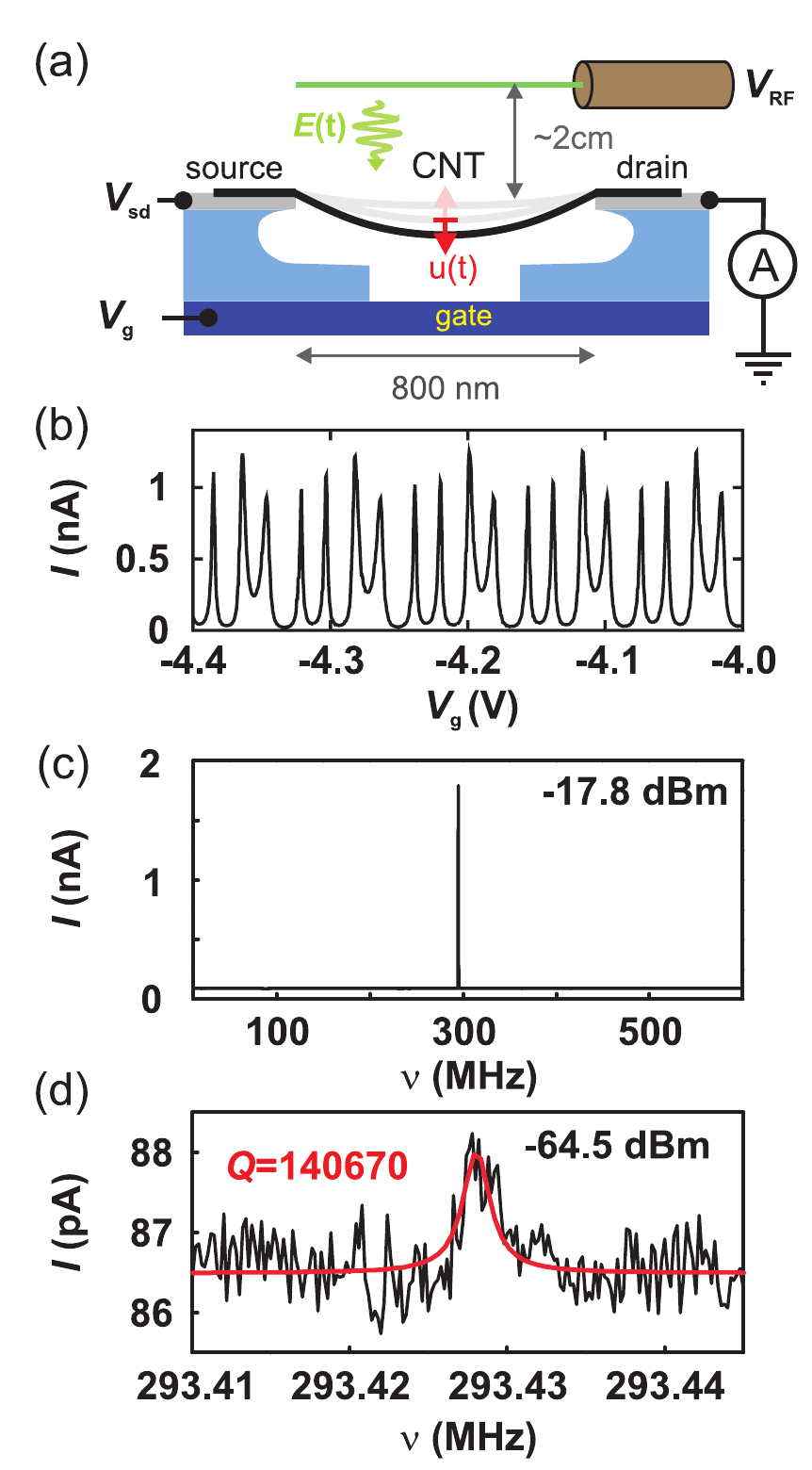, width=8cm}
  \caption{
    \label{fig-intro}
    (a) Schematic drawing of the chip geometry, antenna, and measurement
    electronics. The nanotube acts as a doubly clamped beam resonator,
    driven by an electric field $E(t)$. The displacement of the
    nanotube is $u(t)$.
    (b) Example trace of the dc
    current at $\vsd=50\un{\textmu V}$ as a function of gate voltage,
    demonstrating the regularity of the Coulomb peaks. It shows
    the  four-fold degeneracy typical for clean single-wall carbon nanotubes.
    (c) When the frequency $\nu$ of an RF signal
    on the antenna is swept with fixed \vg\ and \vsd, a resonant peak
    emerges in $I(\nu)$. An example of such a resonance is shown for
    a driving power of $-17.8\un{dBm}$ at a temperature of 20 mK.
    (d) Zoom of the resonance of (c) at low power ($-64.5\un{dBm}$).
    The red line is a fit of a squared damped driven harmonic oscillator
    response to the resonance peak. For both (c) and (d)
    $\vg = -5.16\un{V}$ and $\vsd=0.35\un{mV}$.
    }
\end{figure}

Suspended carbon nanotube devices are made by growing nanotubes
between platinum electrodes over an 800 nm wide pre-defined
trench. The device geometry is shown in \ref{fig-intro}(a). The
fabrication method is discussed in detail by Steele {\it et al.}
 \cite{gary-submitted}. There, the device includes three local
gates for tuning the confinement: here, however, we apply the
same voltage to all three gates, so that they act together as one
single gate. Since no device processing takes place after
nanotube growth and the entire device is suspended, the nanotubes
are highly defect-free and do not suffer from potential
irregularities induced by the substrate surface
 \cite{gary-submitted,nmat-cao:745}. The fabrication method also
offers the advantage that the resonator is not contaminated with
resist residues.

After fabrication, the suspended nanotube devices are mounted in
a dilution refrigerator with filtered twisted pair cabling
attached to source, drain, and gate contacts (see
~\ref{fig-intro}(a)). This configuration allows us to apply dc
gate and bias voltages to the suspended nanotube, and measure the
current flowing through it. To minimize heating, we drive the
nanotube resonator with the electric field radiated from a radio
frequency (RF) antenna positioned near the sample ($\sim$~ 2 cm)
instead of connecting high-frequency cables directly to the
sample. Measurements are performed at temperatures down to the
base temperature of the mixing chamber of the dilution
refrigerator, $T_\text{MC}\simeq 20\un{mK}$.

\ref{fig-intro}(b) shows the Coulomb oscillations of a
semiconducting carbon nanotube with a suspended length of 800~nm.
A highly regular addition spectrum with clear four-fold
degeneracy is visible, characteristic for a defect-free
single-wall carbon nanotube  \cite{prl-oreg:365,prl-liang:126801}.
From the magnetic-field dependence of the position of the Coulomb
oscillations close to the semiconducting
gap \cite{nature-minot:536}, we find the radius of the nanotube,
$r$, to be between 1-1.5 nm. The value of the semiconducting gap
$\approx 0.3 \un{eV}$ is estimated from the gate range between
electron and hole conduction in the device at low temperatures,
which is in agreement with this value for $r$.

When an ac voltage $\vaac\ $ with frequency $\nu$ is applied to
the antenna, we observe a resonant feature at a well-defined
frequency in the dc current flowing through the nanotube.
~\ref{fig-intro}(c) shows an example of such a measurement at a
large radio frequency (RF) voltage, or equivalently a high
generator power. A sharp resonant feature is clearly visible at
$\nu = 294\un{MHz}$. Zooming in on this feature at a lower power
(\ref{fig-intro}(d)) reveals a resonance peak with a narrow
lineshape. A numerical fit of this data yields a quality factor
$Q=140670$ (see below for a discussion of the expected
lineshape). We have also performed measurements on a second
device displaying similar resonant peaks with Q-factors up to
20000; the results on that device are shown in the Suppl.
Information.

\begin{figure}[ht]
\centering
    \epsfig{file=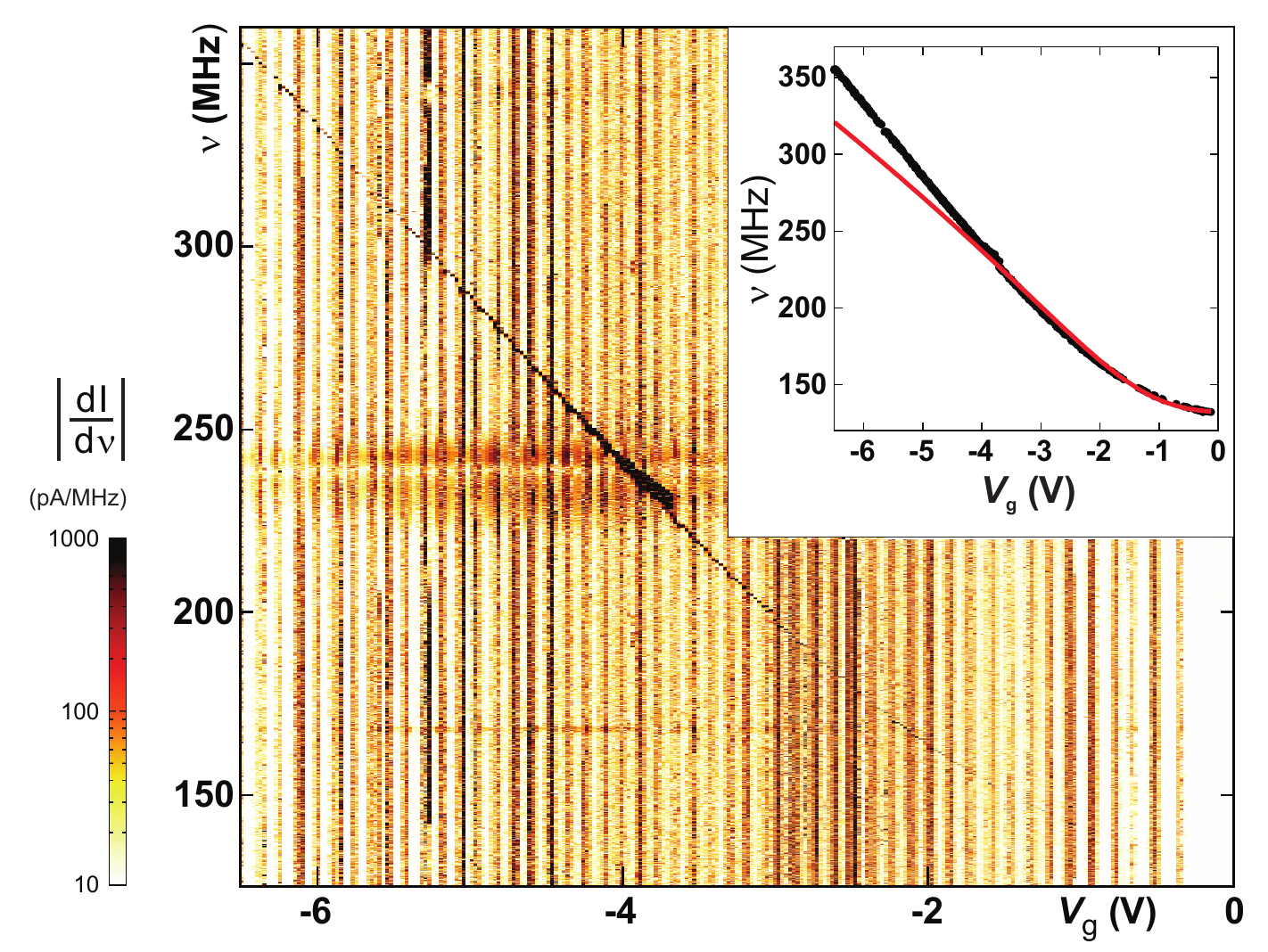, width=14cm}
    \caption{\label{fig-gatedep} $\left| dI/d\nu \right|$
    as a function of frequency $\nu$ of the ac voltage on the antenna
    and the dc gate voltage $\vg\ $ on the back-gate electrode.
    Horizontal stripes are caused by electrical (cable)
    resonances \cite{nature-sazonova:284,nl-witkamp:2904}; the narrow
    vertical stripe pattern is related to the Coulomb blockade oscillations.
    In addition, a gate-dependent resonant feature is clearly visible. Inset:
    Comparison of the extracted resonance frequency to the continuum
    model for the bending mode with $\nu_{\text{bending}} = 132.0$ MHz, $\vgstar =
    2.26 \un{V}$, $T_0 = 0$, and a shift of $0.775\un{V}$ in gate voltage to account
    for an offset in the charge neutrality point of the nanotube from $\vg = 0 \un{V}$
    and the band gap region \cite{nl-witkamp:2904,pssb-poot:4252}. The parameters are
    discussed in the text. An apparent shift of the mechanical resonance frequency at
    $\nu_0 \simeq 230\un{MHz}$ is caused by an electrical (cable) transmission resonance,
    leading to a strong increase in transmitted RF power and distorted peak shapes.
    }
\end{figure}

The resonance observed in parts (c) and (d) of \ref{fig-intro} can be
attributed to the flexural vibration mode of the suspended
nanotube  \cite{nl-chiu:0, nl-lassagne:3735, nature-sazonova:284,
nl-witkamp:2904}. To verify this, we electrostatically induce
tension in the nanotube by applying a dc gate voltage $\vg\ $ to
the back gate electrode
 \cite{nature-sazonova:284,nl-witkamp:2904}. The dc gate voltage
dependence is shown in \ref{fig-gatedep}. When decreasing the
gate voltage from zero to more and more negative values, the
resonance is tuned to higher frequencies by almost a factor of
three: from less than $\nu_0 = 140 \un{MHz}$ at $\vg = -1 \un{V}$
to $\nu_0 = 355.5 \un{MHz}$ at $\vg\ = -6.5 \un{V}$. For the
latter resonance frequency, the thermal
occupation \cite{pt_schwabroukes} $n = 1/2 + [\exp(h \nu_0/ \kb
T_{MC}) - 1]^{-1}$ would be 1.2 at $20 \un{mK}$, suggesting that
the resonator would be close to its quantum ground state in the
absence of the driving fields required for our detection scheme.

We have extracted the resonance peak positions from the data in
\ref{fig-gatedep} and plotted them in the inset. The red line
shows the gate dependence of the resonance frequency calculated
with a continuum model for the fundamental flexural bending
mode \cite{pssb-poot:4252,nl-witkamp:2904,comment:actension}. The
parameters are $\nu_{\text{bending}} = 132.0$~MHz, $\vgstar =-
2.26V$, and $T_{0} = 0$, where $\nu_{\text{bending}}$ is the
resonance frequency in absence of residual tension $T_{0}$ and
$\vgstar$ marks the cross over between the weak and strong
bending regime. At high gate-voltage the model calculation
deviates slightly from the experimental values. This is so far
not fully understood and may be related to large static
displacements of the nanotube in a complex electrostatic
environment \cite{apl-kozinsky:253101}. The value
$\nu_{\text{bending}} =22.4/2 \pi L^2\cdot r\sqrt{E/\rho}=
132.0$~MHz, assuming a tube length $L=800$~nm, yields a nanotube
radius of 1.6 nm, in good agreement with the band-gap and
magnetic field estimates.

\begin{figure}[ht!]
    \centering
    \epsfig{file=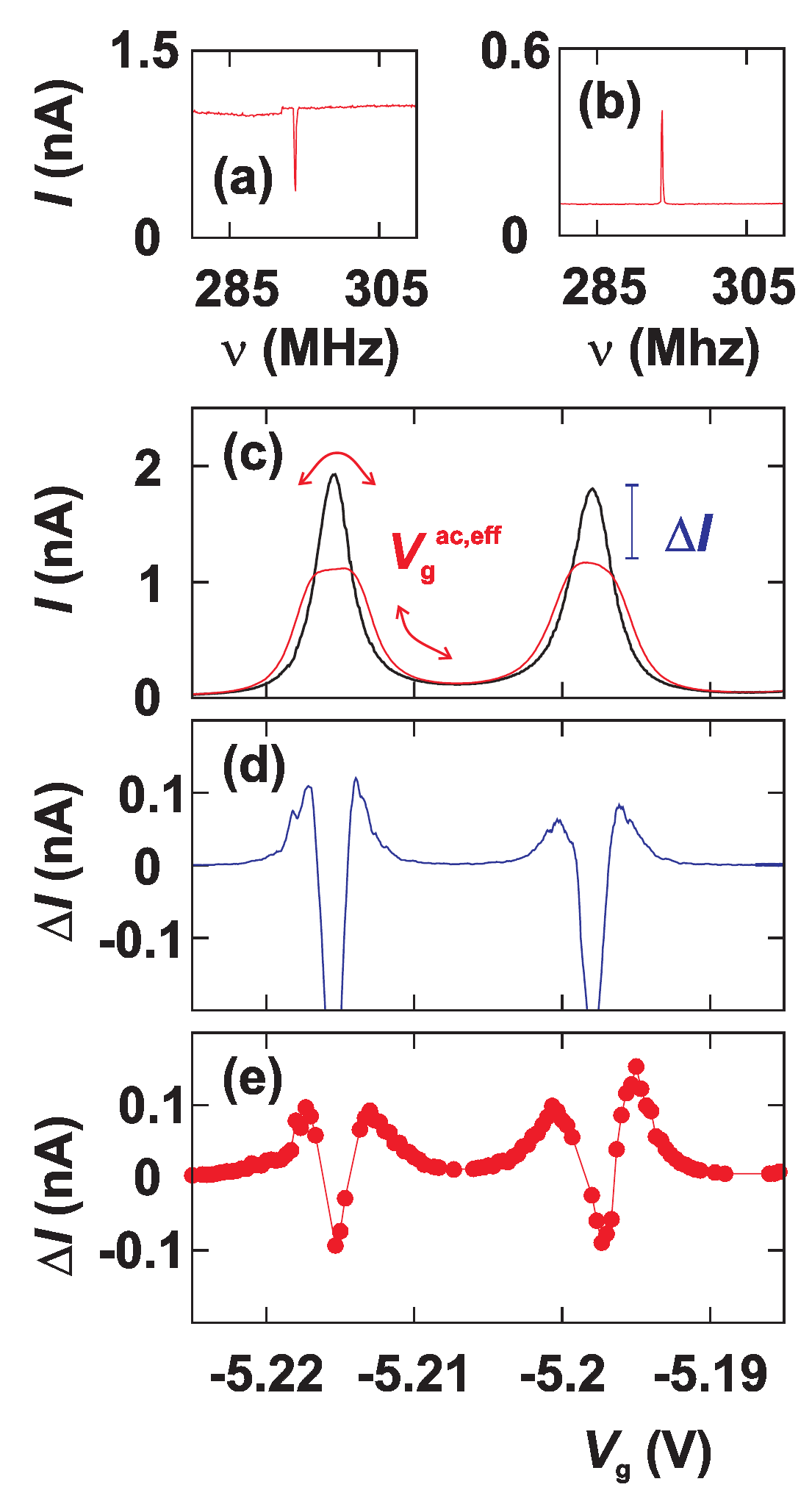, width=7cm}
    \caption{ \label{fig-model} Averaging model for the current
    at resonance. (a), (b) Measured frequency sweeps demonstrating
    the sign change of the resonance amplitude depending on the gate voltage
    (RF power $-13\un{dBm}$, $\vsd=0.1\un{mV}$, $\vg = -5.17\un{V}$ (a) and
    $\vg = -5.16\un{V}$ (b)).
    (c) The black line shows the measured dc current as function of
    gate voltage $I(\vg)$ for $\vsd=0.1\un{mV}$ (no RF). The red line,
    shows the effect of an (effective) ac gate
    voltage on the dc current. This average current (Eq. (\ref{current}))
    is calculated using the measured data and $\vgaceff = 2 \un{mV}$.
    (d) Predicted resonance signal amplitude $\Delta I$ calculated by
    subtracting the dc current from the current averaged over an effective
    gate voltage $\vgaceff = 1 \un{mV}$. For a small $\vgaceff$,
    the signal is proportional to the second derivative $\partial^2 I /
    \partial \vg ^2$ of the black trace shown in (c), as expressed in
    Eq. (\ref{current}). At the top of the Coulomb peak, $\Delta I$ is
    negative, whereas on the flanks of the Coulomb peak, it is positive.
    Note that in (c) a larger value of $\vgaceff$ was used to
    exaggerate the difference between the black and red curves
    for illustrative purposes.
    (e) The measured resonance peak amplitudes obtained from $I(\nu)$
    traces similar to \ref{fig-intro}(d), for $\vsd=0.1\un{mV}$ and a
    RF power of $-48\un{dBm}$.}
\end{figure}

Depending on the gate voltage, the resonance either appears as a
dip (\ref{fig-model}(a)) or as a peak (\ref{fig-model}(b)). Dips
are found around the maxima of the Coulomb oscillations; away
from these maxima peaks are observed. This indicates that the
detection of the mechanical modes is due to electrostatic
interactions as we will now show. We model the effect of a small
change in gate voltage $\dvg $ on the current flowing through the
nanotube by a Taylor expansion of $I(\vg + \dvg)$ around
$\dvg=0$. A crucial point in this expansion is that the second
(and third) order term cannot be neglected, since the current
flowing through the nanotube is strongly non-linear in the
vicinity of the Coulomb oscillations. This is in contrast to the
mixing technique  \cite{nature-sazonova:284,nl-witkamp:2904},
where only the linear term in the expansion is needed.

The motion of the nanotube enters the measured current as
follows: On resonance, the nanotube position $u(t) = u_0 \cos
\left(2\pi \nu_0 t\right)$ oscillates with a finite amplitude
$u_0$, which periodically modulates the gate capacitance $\cg\ $
by an amount $\cgac = (\text{d}\cg / \text{d}u) \; u_0$. The
current flowing through the nanotube does not just depend on the
gate voltage itself; more specifically, it depends on the product
of the gate voltage and the gate capacitance - the so-called
gate-induced charge \cite{nature-sazonova:284, nl-witkamp:2904,
nato-kouwenhoven}. A modulation of the capacitance due to the
motion of the nanotube therefore has the same effect on the
current as if an effective ac gate voltage $\vgaceff = \vg \cgac
/ \cg$ were applied to the gate-electrode. The time-dependent
current can then be calculated by inserting $\dvg = \vgaceff
\cos(2\pi\nu t)$ into the Taylor expansion of $I(\vg + \dvg)$.

Since the mechanical resonance frequency is much larger than the
measurement bandwidth, time-averaged currents are detected in our
setup. We find that the time-averaged mechanically induced current equals:
\begin{equation} \label{current}
\overline{I}(u_0, \vg)  = I(\vg) + \frac{u_0^2}{4}\,
\left(\frac{\vg}{\cg} \frac{\text{d}\cg}{\text{d}u} \right)^2
\frac{\partial^2 I}{\partial \vg ^2} + \mathcal{O}\left( u_0^{4}
\right),
\end{equation}
where only even powers of $u_0$ enter the low-frequency current
due to averaging. The change in dc current on mechanical
resonance $\Delta I = \overline{I} - I$ is thus proportional to
the local curvature $\partial^2 I/\partial \vg^2$ of the Coulomb
blockade oscillations $I(\vg)$.

Using measured Coulomb oscillation traces where no driving signal
was applied (black line in \ref{fig-model}(c)), we have
numerically calculated the behavior of a current time-averaged
over \vgaceff. The result is shown as a red line in
\ref{fig-model}(c). \ref{fig-model}(d) shows the difference
$\Delta I$ between the time-averaged and the static current of
\ref{fig-model}(c). On top of a Coulomb oscillation, the
curvature is negative and the averaged current (red) is smaller
than the static current (black), resulting in a dip in the
current on resonance, when mechanical motion takes place. On the
other hand, $\Delta I$ is positive on the flanks of the Coulomb
oscillations as the curvature is positive there. This can be
compared with the traces $I(\nu)$ shown in \ref{fig-model}(a) and
(b), and with the measurements of $\Delta I$ shown in
\ref{fig-model}(e). Here we plot the amplitude $\Delta I(\nu_0)$
of the mechanical response in the dc current $I(\nu)$ for
different gate voltages. The gate voltage dependence of the
extracted amplitude values in dc current is in good qualitative
agreement with the predictions of the model as shown in
\ref{fig-model}(d).

\begin{figure}[ht!]
    \centering
    \epsfig{file=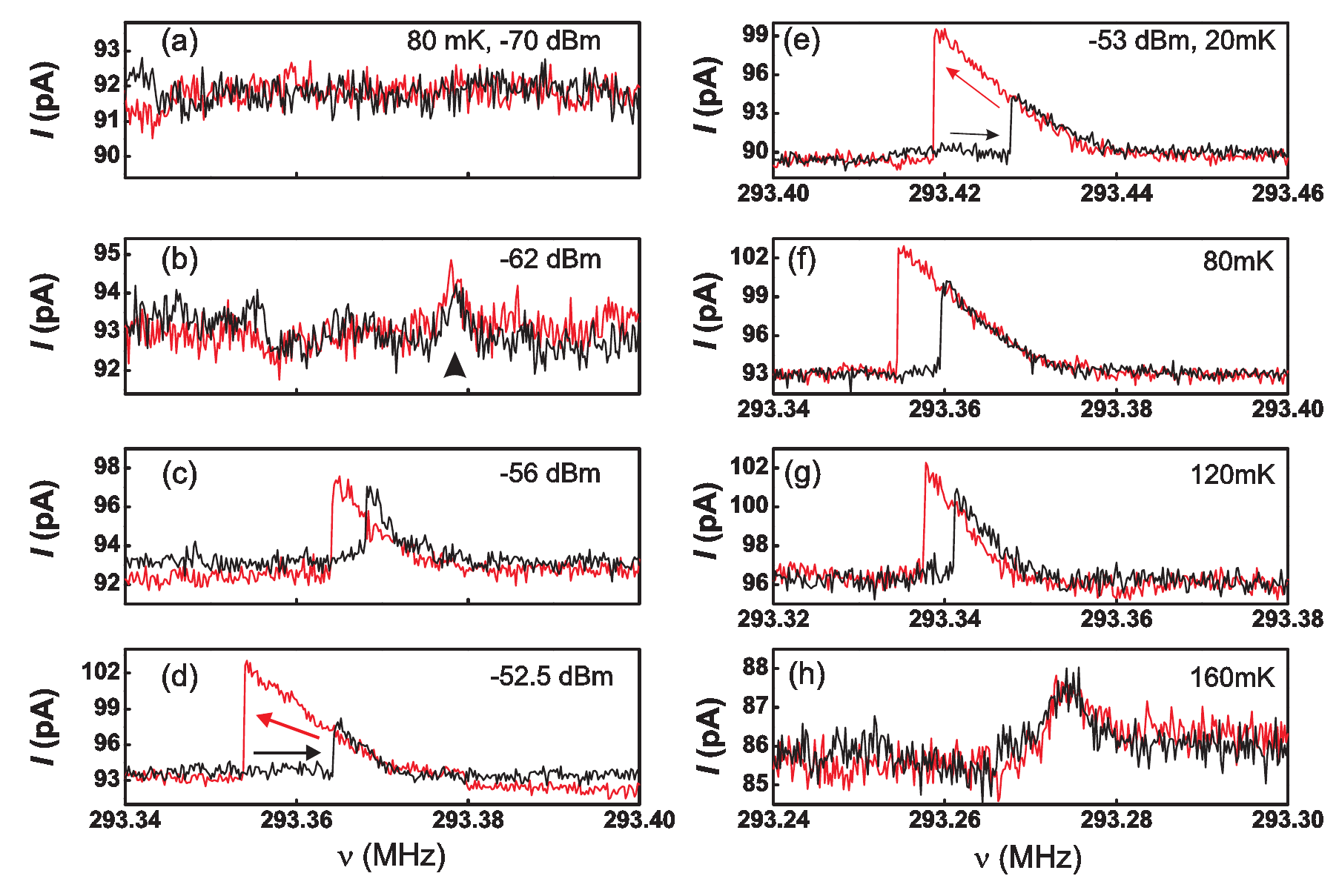, width=14cm}
    \caption{\label{fig-nonlinear} Evolution of the resonance peak
    with increasing driving power (a-d) and temperature (e-h). Black (red) traces
    are upward (downward) frequency sweeps. (a) At low powers, the peak is
    not visible. (b) Upon increasing power, a resonance peak
    with $Q$=128627 appears. (c,d) As the power is increased
    further, the lineshape of the resonance takes on a non-linear
    oscillator form, with a long high frequency tail and a sharp edge at
    lower frequencies. It also exhibits hysteresis between the
    upward and downward sweep that increases with
    driving power, characteristic of a non-linear oscillator. The
    traces (a-d) are taken at 80 mK. (e-h) Forward (black) and reverse
    (red) frequency sweeps at a fixed driving power as a function of
    temperature. At low temperatures,
    the peak shape is non-linear and strongly hysteretic. At the same power,
    but higher temperature, the amount of hysteresis decreases significantly.
    At a temperature of 160 mK, hysteresis and asymmetry are no longer
    apparent; at the same time, the signal amplitude (and with it also signal to
    noise ratio) is decreased, suggesting a decrease in the
    Q-factor with increasing temperature. The working point of traces (a-h) is
    at $\vg = -5.16\un{V}$ and $\vsd=0.35\un{mV}$.}
\end{figure}

The model also allows for a quantitative analysis of the peak
shape and for an estimate of the displacement amplitude $u_0$ in the case of
resonant driving, by evaluating the change in dc current $\Delta I$ alone. We
first note that $u_0$ can be described by the response of a
damped driven harmonic oscillator  \cite{nl-witkamp:2904,
pssb-poot:4252}. From Eq.~\ref{current}, we see that $\Delta I
\propto u_0^2$ so that the measured mechanical response (dip or
peak) in the current is given by the square of the harmonic oscillator response
function (SHO). For the resonance presented in \ref{fig-intro}(d), we find
$\nu_0=293.428\un{MHz}$ and $Q=140670$. This Q-value is nearly
two orders of magnitude higher than previous reported values of
the flexural vibration modes in nanotubes \cite{nl-lassagne:3735,
nl-chiu:0, nature-sazonova:284, nl-witkamp:2904}. Such high
Q-values make this type of device very suitable for mass
detection. From the measured response in \ref{fig-nonlinear}(e)
we estimate (see Suppl. Information) a mass sensitivity of $7
\un{yg/\sqrt{Hz}}$, i.e., in one second it should be possible to
determine if, for example, a He atom has adsorbed onto the
nanotube.

The displacement amplitude $u_0$ in the case of resonant driving
is estimated by modelling the capacitance between the nanotube
and the back-gate as the capacitance between an infinite wire and
an infinite conductive plane  \cite{pssb-poot:4252}. Using a
device length of $L=800$ nm, a tube radius $r=$1.5 nm and a gate
distance $h_0=230$ nm, we obtain $\cg=7.8$ aF and $\text{d} \cg /
\text{d} u = -5.9 \un{zF/nm}$. The calculated capacitance value
is consistent with the experimentally determined value of
$\cg=8.9 \un{aF}$ as determined from the Coulomb peak spacing.
For the resonance in \ref{fig-nonlinear}(b), with $\partial^2 I
/\partial \vg ^2 = 4.43 \mu $A/$V^2$ and $\Delta I(\nu_0) = $1.05
pA, we estimate the oscillation amplitude of the nanotube to be
$u_0(\nu_0) = 0.25\un{nm}$ on resonance. This amplitude is two
orders of magnitude larger than that of the thermal fluctuations
($\kb T/2=m (2\pi\nu_{0})^{2}u_{th}^{2}/2$) of the nanotube
 \cite{book-cleland}, which is $\sim 6.5 \un{pm}$ at 80 mK, and
its estimated zero-point motion  \cite{pt_schwabroukes,
M.D.LaHaye04022004} of 1.9 pm at this gate voltage.

When driving the nanotube resonator with large antenna voltages,
we consistently observe hysteretic peak shapes and a strong
frequency pulling of the resonance peaks (i.e. the frequency
decreases for a larger motion amplitude \cite{book-lifshitz,
prl-cross}).
\ref{fig-nonlinear}(a)-(d) show examples of the shape of the
resonance peak at $\vg = -5.16\un{V}$ and $\vsd=0.35\un{mV}$ for
four different driving powers. Black lines indicate the sweep
direction with increasing frequency; red lines the one with
decreasing frequency. At the lowest power, the mechanical
resonance peak is not visible in the noise. With increasing
driving power the resonance peak first shows a linear response
with its characteristic SHO shape (\ref{fig-nonlinear}(b)). At
higher powers hysteresis sets in, which becomes more pronounced
with increasing RF power. This bistability is consistent with
what is expected for a non-linear mechanical (Duffing)
resonator \cite{book-cleland, book-lifshitz}.

We have studied the dynamic range  \cite{apl-postma:223105,
rsci-ekinci:061101, apl-kozinsky:253101, comment:dynamicrange} in more detail and
found that the driving
powers where the (linear) peak disappears in the noise and where
nonlinearity sets in depend on the temperature. An example of
this effect is shown in \ref{fig-nonlinear}(e)-(h). These panels
show that for a fixed gate voltage and driving power, the
nanotube resonator response changes from non-linear to linear
when the operating temperature is increased from 20~mK to 160~mK.
This temperature-dependent behavior hints at a decrease in
Q-factor as the temperature is increased.

\begin{figure}[ht]
\centering
    \epsfig{file=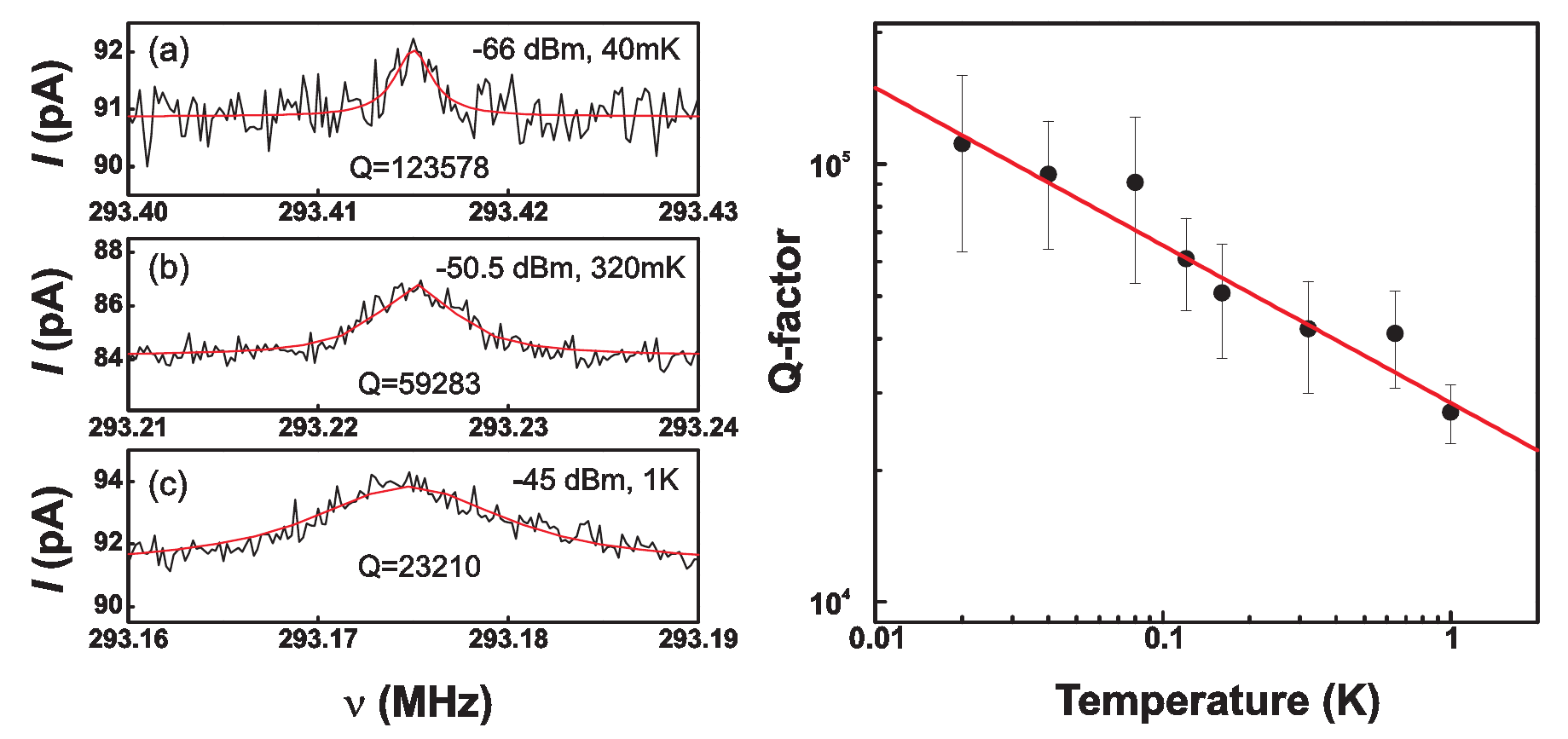, width=17cm}
    \caption{\label{fig-temperatures} Temperature dependence of the
    Q-factor. (a-c) Fits of a squared harmonic oscillator response
    to the resonance in the linear regime at low powers for different
    temperatures at $\vg = -5.16\un{V}$ and $\vsd=0.35\un{mV}$.
    (d) A plot of the Q-factor vs. temperature obtained from linear
    response traces. $Q$ decreases with increasing temperature.
    The red line shows a $T^{-0.36}$ power law dependence (see text).}
\end{figure}

To study the temperature dependence of the quality factor in more
detail, we have determined $Q$ at different temperatures. For a
gate voltage of -5.16~V, three examples of resonance traces are
depicted in \ref{fig-temperatures}(a)-(c). Note that because the
dynamic range is temperature dependent, the RF power is adjusted
at every temperature to ensure a linear response. In
\ref{fig-temperatures}(d), we plot the Q-factor extracted in the
linear regime for eight different temperatures in the range
20~mK$<T_{MC}<1$~K. The error margins are estimated from
ensembles of responses at the same temperature. The Q-factor
changes by a factor four in this temperature range. At the lowest
temperatures, the Q-factor reproducibly reaches values above
10$^5$. These lowest temperature values are close to the
intrinsic Q-values calculated with molecular dynamics simulations
on single-walled carbon nanotube oscillators
 \cite{PhysRevLett.93.185501}. Interestingly, these calculations
predict a $T^{-0.36}$ power law dependence of the Q-factor with
temperature. The red line in \ref{fig-temperatures}(d) shows this
dependence; the data is consistent with this prediction. This
$T^{-0.36}$ dependence has also been observed at low temperatures
in top-down fabricated devices  \cite{shim:133505,
zolfagharkhani:224101}. Note that the Q-values of our nanotube
resonator are much higher than the ones following the trend of
the volume surface ratio in top-down fabricated devices
 \cite{rsci-ekinci:061101}.

In conclusion, using a novel detection mechanism, we have
measured the bending mode resonance of suspended carbon nanotubes
in the single-electron tunneling regime. Sharp gate-tunable
resonances are found with high Q-values ($Q > 10^5$), which can
easily be driven into the nonlinear regime by increasing the
driving power on the RF antenna. By inducing tension with a gate
voltage the frequency can be tuned above $350 \un{MHz}$, so that
the thermal occupation of the resonator approaches 1. Shorter
devices should have even higher resonance frequencies
corresponding to temperatures higher than the mixing chamber
temperature of the dilution refrigerator. These resonators are
therefore in their quantum mechanical ground state, which opens
up the way to new exciting experiments on the quantum aspects of
mechanical motion.

{\bf Acknowledgement.} The authors thank Yaroslav Blanter and Giorgi
Labadze for discussions and Raymond Schouten for experimental
help. This research was carried out with financial support from
the Dutch Foundation for Fundamental Research on Matter (FOM),
The Netherlands Organisation for Scientific Research (NWO),
NanoNed, and the Japan Science and Technology Agency
International Cooperative Research Project (JST-ICORP).

{\bf Supporting Information Available:} Estimation of the mass
sensitivity and measurements on a second device. This material is
available free of charge via the Internet at
{\tt http://pubs.acs.org}.

\bibliography{paper}

\begin{thebibliography}{10}
\bibitem{comment:actension}
An extended continuum model that includes ac tension effects on the gate-tuning
  of the resonance, will be published elsewhere.
\bibitem{comment:dynamicrange}
The lower boundary of the dynamic range is the power at which the resonance is
  no longer visible in the noise. The current noise in our experiments is
  caused by fluctuations in the electrostatic environment of the nanotube and
  not by to the thermal motion of the nanotube.
\end{thebibliography}

\end{document}


\begin{center}

{\LARGE Carbon nanotubes as ultra-high quality factor mechanical
  resonators: Supplementary Information} 

\vspace{2em}

A. K. H\"{u}ttel$^{1,*}$\symbolfootnote[0]{
$^*$Present address: Institute for Experimental
  and Applied Physics, University of Regensburg, 93040 Regensburg,
  Germany},
G. A. Steele$^1$, 
B. Witkamp$^1$, M. Poot$^1$,
L. P. Kouwenhoven$^1$ and H. S. J. van der Zant$^1$ 

{\em $^1$Kavli Institute of NanoScience, Delft University of Technology, PO Box 
5046, 2600 GA, Delft, The Netherlands.}

\end{center}

\section{Mass sensitivity}
The mass sensitivity is estimated from the data plotted in Figure
4(e) of the main text. In the experiments, the mass sensitivity
is limited by the current noise, which has a spectral density
$S_I^{1/2} = 0.12 \un{pA/\sqrt{Hz}}$. The mass sensitivity
$S_m^{1/2}$ can then be calculated as follows: an added mass
$\delta m$ on the nanotube changes the resonance frequency by:
\begin{equation}
\delta \nu_0 = \pder{\nu_0}{m} \delta m = \frac{\nu_0}{2m} \delta
m,
\end{equation}
where $m = 5.1 \times 10^{-21} \un{kg}$ is the mass of a 800 nm
long single-walled nanotube with a 1.5 nm radius. When the
resonance frequency shifts, the current through the nanotube is
modified by:
\begin{equation}
\delta I = \pder{I}{\nu_0} \delta \nu_0 \simeq - \pder{I}{\nu}
\delta \nu_0.
\end{equation}
The latter approximation, which is valid for a high $Q$
resonator, allows us to relate the change in current to the
measured slope of the response function. For the data in Fig.
4(e) the slope of the red line, just right of the jump is
$\partial I / \partial \nu = 6.0 \times 10^{-16} \un{A/Hz}$. The
mass sensitivity is then calculated from $S_m^{1/2} =
\pder{m}{\nu_0} \left ( \pder{I}{\nu} \right )^{-1} S_I^{1/2}$,
which yields $S_m^{1/2} = 7.0 \un{yg / \sqrt{Hz}} = 4.2 \un{u /
\sqrt{Hz}}$. Here, u is the (unified) atomic mass unit, so it is
possible to detect a mass change as small as a single helium atom
within one second.

\vfill

\section{Device B}
\begin{figure}[h]
\begin{center}
    \epsfig{file=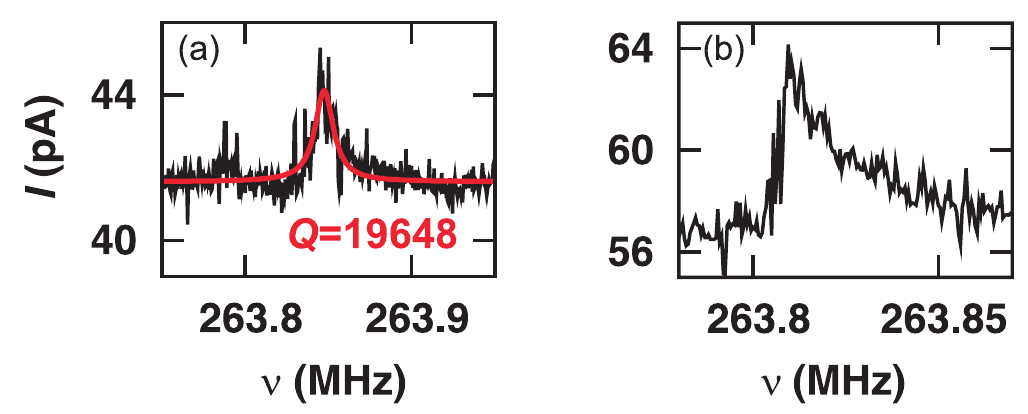, width=12cm}
\end{center}
    \caption{
     Examples of measured resonances in device B in the linear (a)
     and non-linear regime (b) at $20 \un{mK}$.
     Settings: $\vg=-4.241\un{V}$, $\vsd=2.0\un{mV}$, RF power $-47\un{dBm}$ in (a)
     and $\vg=-4.241\un{V}$, $\vsd=1.5\un{mV}$, RF power $-44.5\un{dBm}$ in (b).
     \label{fig:devbpeak}
    }
\end{figure}

\begin{figure}[h]
\begin{center}
\epsfig{file=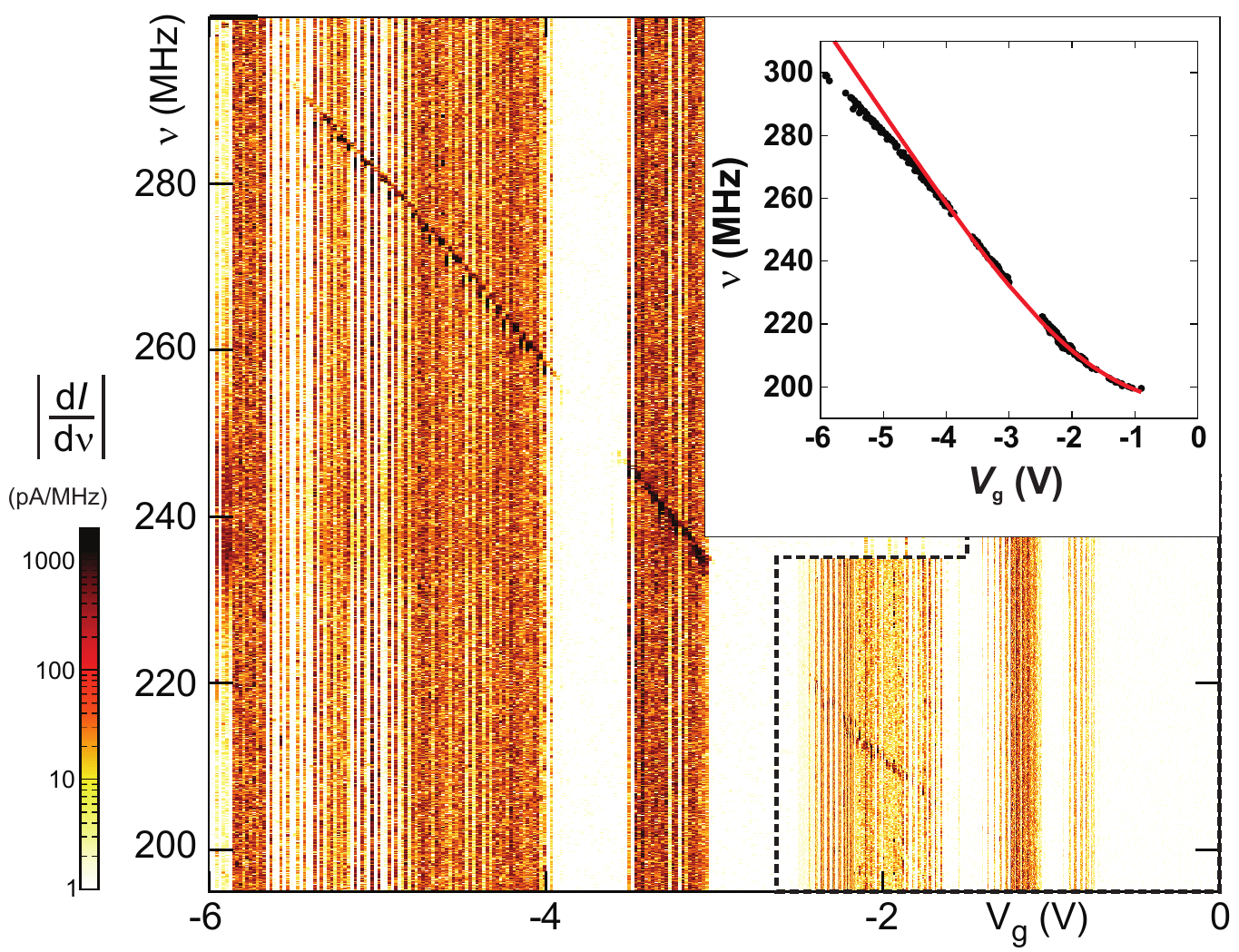, width=14cm} 
\end{center}
\caption{
    $\left| dI/d\nu \right|$ as a function of frequency $\nu$ of the ac
    voltage on the antenna and the dc gate voltage $\vg\ $ on the back-gate
    electrode for device B.
    Left of the dashed line a source-drain voltage of $4 \un{mV}$ was
    used; on the right side $\vsd = 10 \un{mV}$. The RF power was
    $-13\un{dBm}$ everywhere.
    Inset:
    Comparison of the extracted resonance frequency to the continuum
    model for the bending mode with $\nu_{\text{bending}} = 193.7 \un{MHz}$, $\vgstar =
    4.14 \un{V}$, $T_0 = 0$ and a horizontal offset of $1.65 \un{V}$ to account
    for a shift in the charge neutrality point and the band gap region.
    \label{fig:devbgatedep}}
\end{figure}

\ref{fig:devbpeak} and \ref{fig:devbgatedep} show measurements on
a second device. This device also has a suspended length of $800
\un{nm}$. From room temperature measurements it is inferred that
device B is a large ($E_g / k_b > 300 \un{K}$) bandgap nanotube
as well.

